\documentclass[prl,twocolumn,showpacs,preprintnumbers,amsmath,amssymb,mathptmx,citeautoscript,aps,prl]{revtex4-1}    

\usepackage{graphicx}
\usepackage{dcolumn}
\usepackage{bm}
\usepackage{color}
\usepackage{ulem}
\usepackage{cancel}
\usepackage{enumerate}

\begin{document}

\title{Modification of magnetic fluctuations by interfacial interactions in artificially engineered heavy-fermion superlattices}

\author{Genki Nakamine$^1$}
\author{Takayoshi Yamanaka$^1$}
\author{Shunsaku Kitagawa$^1$}
\author{Masahiro Naritsuka$^1$}
\author{Tomohiro Ishii$^1$}
\author{Takasada Shibauchi$^2$}
\author{Takahito Terashima$^1$}
\author{Yuichi Kasahara$^1$}
\author{Yuji Matsuda$^1$}
\author{Kenji Ishida$^1$}

\affiliation{
$^1$ Department of Physics, Kyoto University, Kyoto 606-8502, Japan \\
$^2$ Department of Advanced Materials Science, The University of Tokyo, Kashiwa 277-8561, Japan
}

\date{\today}

\begin{abstract}
Recent progress in the fabrication techniques of superlattices (SLs) has made it possible to sandwich several-layer thick block layers (BLs) of heavy-fermion superconductor CeCoIn$_5$ between conventional-metal YbCoIn$_5$ BLs or spin-density-wave metal CeRhIn$_5$ BLs of a similar thickness.  
However, the magnetic state in each BL, particularly at the interface, is not yet understood, as experimental techniques applicable to the SL system are limited. 
Here, we report measurements of $^{59}$Co-nuclear magnetic resonance, which is a microscopic probe of the magnetic properties inside the target BLs.
In the CeCoIn$_5$/YbCoIn$_5$ SL, the low-temperature magnetic fluctuations of the CeCoIn$_5$ BL are weakened as expected from the Rashba spin-orbit effect.
However, in the CeCoIn$_5$/CeRhIn$_5$ SL, the fluctuations show an anomalous  enhancement below 6 K, highlighting the importance of the magnetic proximity effect occurring near a magnetic-ordering temperature $T_{\rm N} \sim 3$ K of the CeRhIn$_5$ BL.   
We suggest that the magnetic properties of the BLs can be altered by the interfacial interaction, which is a new route to modify the magnetic properties.    
\end{abstract}

\maketitle
Magnetic fluctuations in strongly correlated electron systems have been intensively studied from experimental and theoretical aspects.  
This is primarily related to the fact that most of unconventional superconductors have been discovered in the verge of the magnetic phase, and possess mainly strong antiferromagnetic (AFM) fluctuations\cite{MathurNature1998, MoriyaRPP2003, MonthouxNature2007, KeimerNature2015, ShibauchiARCMP2014}. 
The heavy-fermion (HF) superconductor CeCoIn$_5$ is one such unconventional superconductor. 
The superconducting (SC) transition temperature $T_c$ of CeCoIn$_5$ is 2.3 K, which is the highest $T_c$ among Ce-based HF superconductors\cite{PetrovicJPCM2001}.
CeCoIn$_5$ possesses an extremely narrow conduction band with a relatively heavy effective mass and shares similarities with high-$T_c$ cuprates, including an unconventional SC gap with $d_{x^2-y^2}$ symmetry and non Fermi-liquid behavior\cite{StockPRL2008, IzawaPRL2001, NakajimaJPSJ2007}. 
In addition, it has been considered that the superconductivity is mediated by AFM fluctuations with quantum critical characteristics\cite{AllanNatPhys2013, ZhouNatPhys2013}.
     
Recently, the molecular beam epitaxy (MBE) technique was employed to synthesize artificial Kondo superlattices (SLs) of alternating layers of HF CeCoIn$_5$/conventional-metal YbCoIn$_5$\cite{HuyJMMM2009} and CeCoIn$_5$/spin-density-wave (SDW)-metal CeRhIn$_5$\cite{KohoriEPJ2000, BaoPRB2000} with a few-unit-cell-layer thickness\cite{ShishidoScience2010, MizukamiNatPhys2011}. 
These artificially engineered materials provide a new platform to study the two-dimensional electronic properties of HF superconductors, the interaction between two different block layers (BLs), and the magnetic properties at the interfaces\cite{ShimozawaRepProgPhys2016}.
Particularly, interactions between superconductivity and bosonic excitations through an atomic interface have received much attention since single-layer FeSe on a SrTiO$_3$ substrate shows an extraordinarily high-$T_c$ due to the coupling with the substrate\cite{LeeNature2014,HuangARCMP2017,RademakerNJP2016}. 
Thus, it is important to study electronic and magnetic properties of BLs and the interfaces between two BLs.   
We consider that nuclear magnetic resonance (NMR) and nuclear quadrupole resonance (NQR) are among the best experimental probes for this purpose, as they can provide spatially resolved microscopic information about the target BLs.
We have investigated the magnetic and SC properties in an epitaxial film of CeCoIn$_5$ and the CeCoIn$_5$/YbCoIn$_5$ SL from the $^{115}$In NMR/NQR\cite{YamanakaPRB2017, YamanakaPRB2015}. 
We reported that nuclear spin-lattice relaxation rate $1/T_1$ on the epitaxial film is almost identical to $1/T_1$ of the bulk single crystal, indicative of the high quality of the epitaxial film HF sample\cite{YamanakaPRB2017}, and that $1/T_1$ on the CeCoIn$_5$/YbCoIn$_5$ SL revealed the strong suppression of the magnetic fluctuations at the interface region\cite{YamanakaPRB2015}.    

The interesting questions in such an SL system are whether the magnetic and SC properties on the CeCoIn$_5$ BL can be modified in the SL sample and, if so, what effects and interactions are important, and how many layers are affected in the BL.
We have investigated these issues from the $^{59}$Co ($I$ = 7/2)-NMR measurement of the CeCoIn$_5$/YbCoIn$_5$ and CeCoIn$_5$/CeRhIn$_5$ SL samples. 
In this paper, we report the low-temperature magnetic fluctuations of the CeCoIn$_5$ BL mainly in the CeCoIn$_5$/CeRhIn$_5$ SL.   
The magnetic fluctuations of the CeCoIn$_5$ BL in the CeCoIn$_5$/CeRhIn$_5$ SL is almost unchanged with those of bulk CeCoIn$_5$ down to 6 K, indicating that the coupling between the CeCoIn$_5$ and CeRhIn$_5$ BLs is small. 
However, the magnetic fluctuations of the CeCoIn$_5$ BL is enhanced below 6 K only at the one or two layers at the interface, indicating that the strong magnetic fluctuations penetrate from the CeRhIn$_5$ BL with approaching to the magnetic ordering temperature $T_{\rm N} \sim 3$ K of the CeRhIn$_5$ BL.
By taking into account the experimental results of the CeCoIn$_5$/YbCoIn$_5$ that the magnetic fluctuations of the CeCoIn$_5$ BL are weakened by the Rashba interaction, we revealed that the magnetic properties are modified mainly at the interface region, suggesting that the interfacial interaction is the most important interaction in the SL compounds.

The SLs of CeCoIn$_5$(5)/YbCoIn$_5$(5) and CeCoIn$_5$(5)/CeRhIn$_5$(5), which were grown by the MBE technique, were stacked alternately along the $c$ axis, where (5) indicates the number of unit-cell-layers of each BL. 
$T_c$ of the CeCoIn$_5$/YbCoIn$_5$, and  CeCoIn$_5$/CeRhIn$_5$ SLs is 1.2 and 1.4 K, respectively. 
We also prepared epitaxial-films of CeCoIn$_5$, YbCoIn$_5$, and CeRhIn$_5$, and performed NMR measurements on these films and single-crystal CeCoIn$_5$ for comparison with the SL samples. 
Details of the synthesis process and sample characterization are shown in the reference\cite{NaritsukaPRL2018, SM}

\begin{figure}
\includegraphics[width=0.8\linewidth]{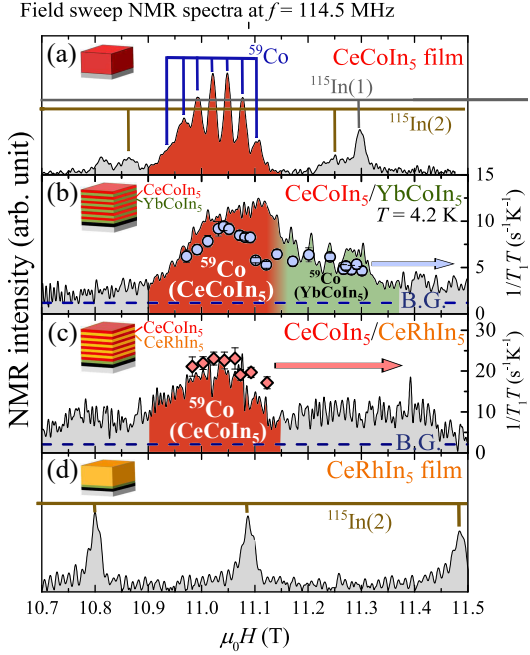}
\caption{Field-sweep nuclear magnetic resonance (NMR) spectra at 4.2 K and 114.5 MHz on (a)  500-nm-thick CeCoIn$_5$ film, (b) CeCoIn$_5$(5)/YbCoIn$_5$(5) superlattice (SL),  (c) CeCoIn$_5$(5)/CeRhIn$_5$(5) SL and (d) 300-nm-thick CeRhIn$_5$ film. The NMR signals in the film samples are identified, and the NMR parameters are listed\cite{SM}.}  
\label{fig1}
\end{figure}
Figure 1 presents the $^{59}$Co and $^{115}$In-NMR spectra of the (a) CeCoIn$_5$ film, (b) CeCoIn$_5$/YbCoIn$_5$ SL, (c) CeCoIn$_5$/CeRnIn$_5$ SL, and (d) CeRhIn$_5$ film, which were measured by sweeping the magnetic field parallel to the $c$-axis with $f$ = 114.5 MHz at 4.2 K. 
In the CeCoIn$_5$ film samples, the NMR signals arising from the Co site have a fine structure because of the presence of the electric field gradient (EFG)\cite{YamanakaPRB2015}.
The parameters of the $^{115}$In and $^{59}$Co NMR spectra are listed in the table\cite{SM}. 
The $^{59}$Co-NMR spectra of the SLs are observed around the field range where the $^{59}$Co-NMR spectra of the film are observed, but they become structureless mainly due to the inhomogeneity of the EFG at the Co site. 
In addition, finite intensity was observed everywhere in the SL spectra, which arises from the In(1) and In(2) sites, since the EFG of these sites are much larger than that of the Co site and the inhomogeneity of the EFG makes the spectra extremely broader.   
The $^{59}$Co-NMR spectrum in the CeCoIn$_5$/CeRnIn$_5$ SL arises only from the CeCoIn$_5$ BL, as no Co atoms exist in the CeRhIn$_5$ BL.

Nuclear spin-lattice relaxation rate $1/T_1$ was measured by the recovery of the nuclear magnetization $m(t)$ at a time $t$ after a saturation pulse. 
The relaxation of $R(t) \equiv [m(\infty)-m(t)]/m(\infty)$ at the $^{59}$Co-NMR peak of the CeCoIn$_5$ BL in two SLs, together with that of the CeCoIn$_5$ film can be fit to a theoretical function $f(t)$ for a spin 7/2 relaxation at the central transition\cite{NarathPR1967}, which can be expressed as 
\begin{eqnarray}
R(t) & \propto & f(t) =  0.012\exp \left(-\frac{t}{T_1}\right)     +0.068 \exp \left(-\frac{6t}{T_1} \right) \nonumber \\
& &  + 0.206\exp \left(-\frac{15t}{T_1}\right)  + 0.714\exp \left(-\frac{28t}{T_1}\right) .
\label{eq_1}
\end{eqnarray}
$1/T_1$ can be evaluated from the best fits. 
The experimental data and the fitting in the CeCoIn$_5$ / YbCoIn$_5$ SL are shown by the curves in Fig. ~S2 \cite{SM}.          

We also measured $1/T_1$ at various fields at 4.2 K in two SLs. 
In the measurement, $1/T_1$ was estimated with the same theoretical function of Eq. (\ref{eq_1}) to compare the $1/T_1$ values, although the tail of the NMR peak arises from the nuclear-spin transitions other than the central transition or even from the back-ground $^{115}$In signals.
The $1/T_1$ obtained with the above procedure is plotted in Figs.~\ref{fig1} (b) and (c), and they show a position dependence. 
It should be noted that the NMR peak near the central transition in the CeCoIn$_5$ film has the largest $1/T_1$. 
This is because the coefficient of $\exp (-28t/T_1)$ in the theoretical function is the largest at the central transition, thus $1/T_1$ should be the largest if the fitting for the evaluation of $1/T_1$ was performed by the same theoretical function.
We can determine the $^{59}$Co-NMR peak arising mainly from the central transition with the $1/T_1$ measurement, even though the NMR spectrum from the CeCoIn$_5$ BL is so broad that it is structureless.
Temperature dependence of $1/T_1$ of the BL was measured at the central-transition peak.
The measurement of $1/T_1$ at the CeCoIn$_5$ BL is crucially important, as we are interested in the comparison between the CeCoIn$_5$/YbCoIn$_5$ SL and CeCoIn$_5$/CeRnIn$_5$ SL.
It is expected that the magnetic fluctuations at the CeCoIn$_5$ BL will be different between the two SLs because of the different adjacent BLs.   

\begin{figure}
\hspace{-0.3cm}
\includegraphics[width=1.0\linewidth]{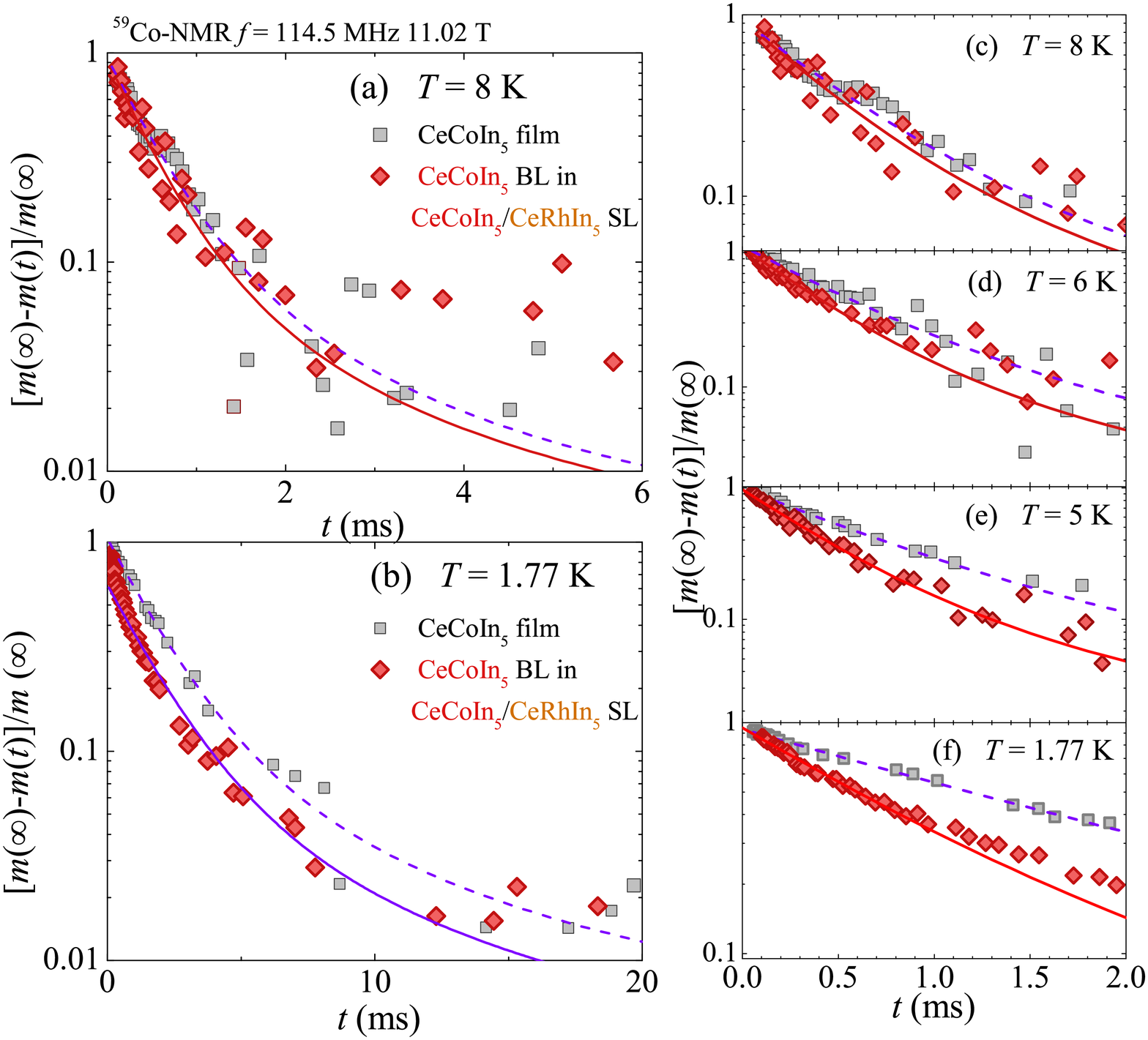}
\caption{Relaxations of $R(t)$ of the CeCoIn$_5$ BL in the CeCoIn$_5$/CeRhIn$_5$ SL and $R(t)$ of the CeCoIn$_5$ film at (a) 8 and (b) 1.77 K. 
(a) At 8 K, two $R(t)$ can be fit with the theoretical function with the almost same $1/T_1$ over the full time range.
(b) At 1.77 K, $R(t)$ of the CeCoIn$_5$ BL in the CeCoIn$_5$/CeRhIn$_5$ SL has a larger $1/T_1$ component at the short time range, and a smaller $1/T_1$ component can be fitted with the same $1/T_1$ component of the CeCoIn$_5$ film in $t > 1$ ms, which is represented by the solid curve in the main figure. The solid curve is the $0.6*f(t)$, and $f(t)$ (the dotted line) is the fitting curve of $R(t)$ of the CeCoIn$_5$ film.    
(c), (d), (e) and (f) are the $R(t)$ in the time range between 0 and 2 msec at $T$ = 8, 6, 5, and 1.77 K, respectively. The larger component of $1/T_1$ can be recognized below 6 K.   }
\label{fig2}
\end{figure}
A clear enhancement in the AFM fluctuations at the interface region was observed in the CeCoIn$_5$/CeRhIn$_5$ SL at low temperatures as discussed below.
Figure \ref{fig2} (a) and (b) show the relaxation of $R(t)$ for the CeCoIn$_5$ BL in the CeCoIn$_5$/CeRhIn$_5$ SL and the CeCoIn$_5$ film at 8 and 1.77 K, respectively. 
Although $R(t)$ for the CeCoIn$_5$ BL in the CeCoIn$_5$/CeRhIn$_5$ SL exhibits the same relaxation behavior as that for the CeCoIn$_5$ film at 8 K, the former $R(t)$ has a larger component of $1/T_1$ (shorter component of $T_1$) than the latter $R(t)$ at 1.77 K.
The larger component of $1/T_1$ can be recognized particularly in the short time range less than 2 ms below 6 K as seen in Fig. \ref{fig2} (c - f), but the smaller component of $1/T_1$ (the longer component of $T_1$) is almost the same as $1/T_1$ in the CeCoIn$_5$ film, as the $R$($t$) smaller than $1/e$ can be fitted to the relaxation curve with the similar $1/T_1$ of the film with the different initial value, as seen in the main panel of Fig.~2 (b).
Thus, the larger component of $1/T_1$ for the CeCoIn$_5$ BL in the CeCoIn$_5$/CeRhIn$_5$ SL was evaluated below 5.5 K from the fit in the short time range where $1 > R(t) > 1/e$, as the larger component of $1/T_1$ can be observed only below 5.5 K, and the smaller component of $1/T_1$ was estimated from the time range where $R(t)$ of CeCoIn$_5$ BL is smaller than $1/e$.                           
\begin{figure}
\includegraphics[width=0.9\linewidth]{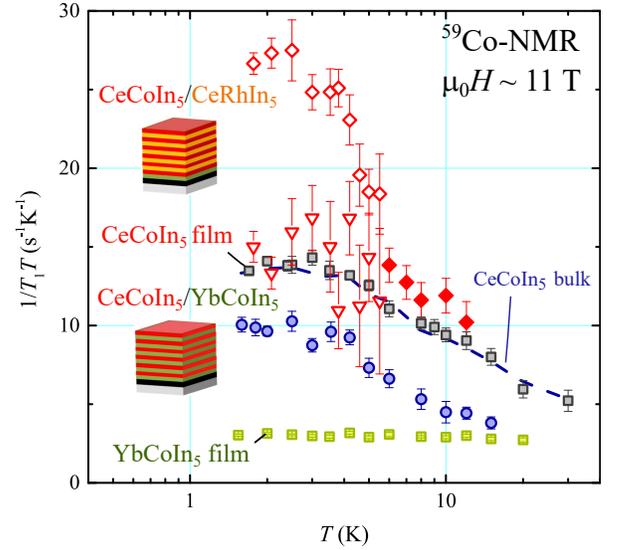}
\caption{Temperature dependence of $1/T_1T$ for $^{59}$Co on the CeCoIn$_5$ BL in the CeCoIn$_5$/YbCoIn$_5$ and CeCoIn$_5$/CeRhIn$_5$ SLs together with the CeCoIn$_5$ and YbCoIn$_5$ films.
As for the $1/T_1T$ of the CeCoIn$_5$/CeRhIn$_5$ SL, the larger (red open diamonds) and smaller (red open triangles) $1/T_1T$ components are plotted below 5.5 K. $1/T_1T$ above 6K was evaluated from the single component shown by red closed diamond.  
The $1/T_1T$ of $^{59}$Co on the single-crystal CeCoIn$_5$ is represented by a dotted line. }
\label{fig3}

\end{figure}

\begin{figure}
\includegraphics[width=1\linewidth]{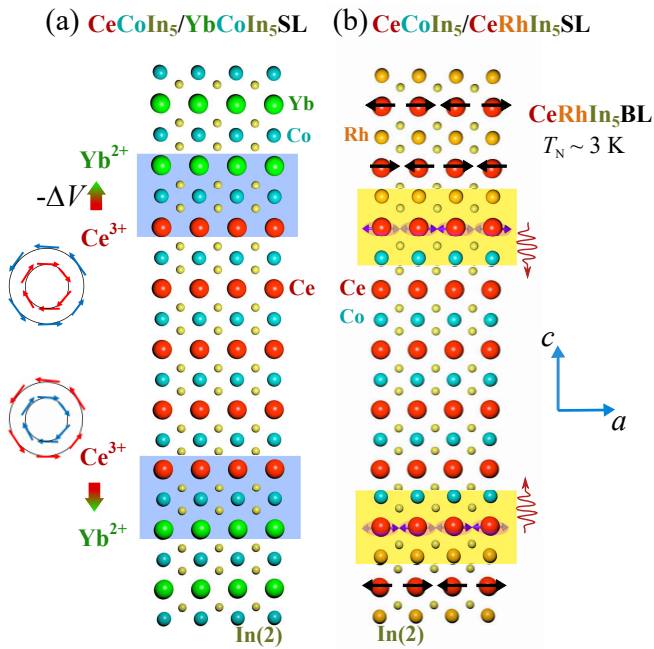}
\caption{Arrangement of the atomic layers revealed with a high-resolution transmission electron microscope image in (a) CeCoIn$_5$/YbCoIn$_5$ and (b) CeCoIn$_5$/CeRhIn$_5$ BLs. 
The atomic views on the (1, 0, 0) plane are shown. 
The interfaces of each BL layers are indicated by boxes, and the conceivable effects are shown.}
\label{fig4}

\end{figure}

Figure \ref{fig3} shows the temperature dependence of $1/T_1T$ for $^{59}$Co on the CeCoIn$_5$ BL in the CeCoIn$_5$/CeRhIn$_5$ and CeCoIn$_5$/YbCoIn$_5$ SLs together with  that of the CeCoIn$_5$ and YbCoIn$_5$ films  and the bulk CeCoIn$_5$. 
As for the $1/T_1T$ of the CeCoIn$_5$/CeRhIn$_5$ SL, the larger and smaller components of $1/T_1$'s are plotted below 5.5 K.
$1/T_1T$ for the CeCoIn$_5$ BL in the CeCoIn$_5$/YbCoIn$_5$ SL, which is smaller than $1/T_1T$ of the CeCoIn$_5$ film, indicates the suppression of the AFM fluctuations in the CeCoIn$_5$ BL.
This result is consistent with our previous $^{115}$In-NMR results for the CeCoIn$_5$/YbCoIn$_5$ SL\cite{YamanakaPRB2015}.
A stronger suppression of the AFM fluctuations at the interface was shown by the site selective NMR measurements\cite{YamanakaPRB2015}, with which we succeeded in identifying the $^{115}$In-NMR signals arising from the interface and the inner layers separately.
However, such a local information could not be obtained from the $^{59}$Co-NMR measurement due to the smallness of the EFG at the Co site.     
In the absence of the local inversion symmetry of the Ce compounds with the strong correlations, the Rashba-type spin-orbit interaction is dominant and splits the Fermi surfaces into two sheets depending on spin structure. 
An image of this effect is presented in Fig.~\ref{fig4} (a), and, as a result, the nesting condition is modified and the commensurate AFM fluctuations with $\mbox{\boldmath $Q$} = (\pi /a, \pi /a, \pi/c)$, which is dominant in CeCoIn$_5$, are markedly suppressed.

In contrast, it should be noted that the $1/T_1T$ for the CeCoIn$_5$ BL in the CeCoIn$_5$/CeRhIn$_5$ SL is nearly the same as that in the bulk CeCoIn$_5$ down to 6 K. 
This indicates that AFM fluctuations at the interface region between the CeCoIn$_5$ BL and CeRhIn$_5$ BL are not suppressed by the Rashba spin-orbit interaction, and that the coupling between the CeCoIn$_5$ and CeRhIn$_5$ BLs is small, as magnetic fluctuations at the CeRhIn$_5$ BL would be larger than that of the CeCoIn$_5$ BL. 
Below 6 K, $1/T_1T$ has a larger component than that of the CeCoIn$_5$ film, although the smaller component of $1/T_1T$ is almost identical to that of the CeCoIn$_5$ film, as mentioned above. 
This indicates that the AFM fluctuations are enhanced because of the magnetic interaction between the two BLs.
The larger component of $1/T_1T$ in the CeCoIn$_5$ BL continues to increase down to 1.77 K, although a kink is observed at about 2.5 K, which is slightly lower than the magnetic ordering temperature ($T_{\rm N} \sim 3$ K) of the CeRhIn$_5$ BL. 
The $1/T_1T$ result indicates that the AFM fluctuations at the CeCoIn$_5$ BL remain enhanced below the $T_{\rm N}$ of the CeRhIn$_5$. 
The paramagnetic state of the CeCoIn$_5$ BL below $T_{\rm N}$ is also shown from the absence of appreciable broadening of the $^{59}$Co-NMR spectrum by the internal field arising from the Ce ordered moment as shown in Fig. ~S3\cite{SM}.
As the smaller component of $1/T_1T$ for the CeCoIn$_5$ BL is similar to that of the CeCoIn$_5$ film and the fraction of this component is roughly estimated to be half of the total relaxation component, it is reasonable to conclude that the enhancement in the AFM fluctuations only occurs at the Ce layer sandwiched by the Co and Rh atoms at the interface or the Ce layers next to this Ce layer as shown in Fig.~\ref{fig4} (b) since there remain the layers where the same AFM fluctuations as those in the CeCoIn$_5$ film persist. 
Our NMR results indicate that the Ce layers at the interface are not in the ordered state, but would remain in the paramagnetic state and become the origin of strong AFM fluctuations in the CeCoIn$_5$ BL.
This magnetic proximity might be regarded as an injection of a strong AFM paramagnon into the CeCoIn$_5$ BL from the adjacent SDW-metal CeRhIn$_5$ through the interface, although this injected region is considered to be only one or two lattices along the $c$-axis as discussed above.  

Up to now, we have not obtained the NMR results in the SC state as our NMR measurement has been done above the upper critical field ($H_{\rm c2}$), but we consider that superconductivity of the CeCoIn$_5$/CeRhIn$_5$ SL would be
 in the stronger coupling nature than that of bulk CeCoIn$_5$ from the recent result of the ratio of $H_{\rm c2} /T_c$, although $T_c$ of the SL is lower\cite{NaritsukaPRL2018}.
This is suggested by the result that superconductivity in the CeCoIn$_5$ BLs possess an extremely strong coupling nature when AFM order in the CeRhIn$_5$ BLs vanishes\cite{NaritsukaPRL2018}.
We suggest that the AFM paramagnons injected through the interface from the CeRhIn$_5$ BLs would work to strengthen the pairing interactions.
This picture is similar to the suggested mechanism of the higher-$T_c$ in the mono-layer FeSe, where pronounced oxygen optical phonons in the substrate couple with the mono-layer FeSe electrons and enhance the superconductivity\cite{Song}.  

In conclusion, from NMR studies on two SLs, we demonstrate that the dominant interaction working at the interface region strongly depends on the characteristics of the adjacent BLs and that the magnetic properties of the CeCoIn$_5$ BL are modified by the penetration of the interfacial magnetic properties into the inner layers. 
We suggest that the interfacial interaction is a key factor to tune the magnetic characteristics of the BL and will become a new method to control the magnetic properties.  
To prove our suggestion, the spectroscopic experiments which can investigate the thickness dependence of magnetic fluctuations are highly desired.
We anticipate that AFM-fluctuation-mediated superconductivity, which is realized only in the interface region, might be induced in an SL consisting of the conventional-metal and SDW-metal BLs.  
 
The authors acknowledge H. Ikeda and Y. Yanase for fruitful discussions. 
This work was partially supported by Kyoto Univ. LTM center, Grant-in-Aid from the Ministry of Education, Culture, Sports, Science, Technology(MEXT) of Japan, Grants-in-Aid for Scientific Research (KAKENHI) from the Japan Society for the Promotion of Science (JSPS), the ``J-Physics'' (No.\,JP15H05882, No.\,JP15H05884, and No.\,JP15K21732) and ``Topological Quantum Phenomena'' (No.\,JP25103713) Grant-in-Aid for Scientific Research on Innovative Areas from the MEXT of Japan, and by Grant-in-Aids for Scientific Research (Grants No. JP25220710, and No. JP15H05745).

%

\end {document}